\documentclass[twocolumn,showpacs,prl]{revtex4}%
\usepackage{amssymb}
\usepackage{graphicx}
\usepackage{dcolumn}
\usepackage{bm}
\usepackage{amsmath}
\usepackage{amsfonts}%
\begin{document}
\title{Bragg scattering of Cooper pairs in an ultra-cold Fermi gas}
\author{K. J. Challis}
\email{kchallis@physics.otago.ac.nz}
\author{R. J. Ballagh}
\author{C. W. Gardiner}
\affiliation{Jack Dodd Centre for Photonics and Ultra-Cold Atoms \\
Department of Physics, University of Otago, P.O. Box 56, Dunedin, New Zealand}
\date{\today}

\begin{abstract}
We present a theoretical treatment of Bragg scattering of a degenerate Fermi
gas in the weakly interacting BCS regime.  Our numerical calculations predict correlated scattering of Cooper pairs into a spherical shell in momentum space. The scattered shell of correlated atoms is centered at half the usual Bragg momentum transfer, and can be clearly distinguished from atoms scattered by the usual single-particle Bragg mechanism.  We develop an analytic model that explains key features of the correlated-pair Bragg scattering, and determine the dependence of this scattering on the initial pair correlations in the gas.
\end{abstract}

\maketitle

Bragg scattering provides a high precision spectroscopic technique that has
been adapted from materials science to probe Bose-Einstein condensates
\cite{Phillips99, Ketterle99}. In condensate systems, signatures of soliton
evolution \cite{Ertmer03}, phase fluctuations \cite{Aspect03},
centre-of-mass motion \cite{Wilson03}, and vortex structure \cite{Raman06},
are accessible due to the velocity selectivity of Bragg spectroscopy. It has
been proposed \cite{Ruostekoski00, Torma02, Zoller04, Deb06, Combescot06} that Bragg
spectroscopy of an ultra-cold Fermi gas can provide insight into the Cooper
paired regime, and the transition through a Feshbach resonance to molecule formation.

In this paper we develop Bragg spectroscopy as a probe of ultra-cold atoms by
investigating Bragg scattering of a weakly attractive Cooper paired Fermi gas.
Our calculations differ from existing theoretical treatments by: (i) providing
explicit solutions for the time evolution of the matter field subjected to a
moving optical grating, allowing direct observation of the dynamic response of
the gas in momentum space, (ii) investigating in detail the large momentum
transfer regime, where the atoms scatter well outside the Fermi sea, and,
(iii) determining the Bragg spectrum of the Fermi gas in an analogous way to
the case for a Bose-Einstein condensate \cite{Ketterle99,Blakie02}, by
calculating the momentum transfer per atom over a range of Bragg frequencies.
The key result we report is Bragg scattering of correlated atom pairs via
generation of a Bragg grating in the pair potential.

Our theoretical treatment is based on a mean-field description of a degenerate
weakly attractive homogeneous Fermi gas.  Two spin states are present in equal numbers, with field operators $\hat{\psi}_{\uparrow}(\mathbf{r},t)$ and $\hat{\psi}_{\downarrow}(\mathbf{r},t)$, and the collisional interaction
Hamiltonian is approximated by a number of single-particle terms.  For convenience, the optical Bragg field is chosen so that it does not flip the particle spin. Implementing the Heisenberg equations of motion and the Bogoliubov transformation,
\begin{equation}
\hat{\psi}_{\uparrow,\downarrow}(\mathbf{r},t)=\sum_{\mathbf{k}} \left[
u_{\mathbf{k}}(\mathbf{r},t)\hat{\gamma}_{\mathbf{k} \uparrow,\mathbf{k}\downarrow} \mp v_{\mathbf{k}}^{*}(\mathbf{r},t)\hat{\gamma}^{\dagger
}_{\mathbf{k} \downarrow,\mathbf{k}\uparrow} \right]  ,
\end{equation}
yields \cite{deGennes,K&S,Challis07} the time-dependent Bogoliubov de Gennes
equations,
\begin{align}
i\hbar\frac{\partial}{\partial t} \left[
\begin{array}
[c]{c}
u_{\mathbf{k}} (\mathbf{r},t)\\
v_{\mathbf{k}}(\mathbf{r},t)
\end{array}
\right]  = \left[
\begin{array}
[c]{cc}
L(\mathbf{r},t) & \Delta(\mathbf{r},t)\\
\Delta^{*}(\mathbf{r},t) & -L(\mathbf{r},t)
\end{array}
\right]  \left[
\begin{array}
[c]{c}
u_{\mathbf{k}}(\mathbf{r},t)\\
v_{\mathbf{k}}(\mathbf{r},t)
\end{array}
\right]  , \label{bdgeqns}%
\end{align}
where
\begin{equation}
L (\mathbf{r},t)=-\frac{\hbar^{2}\nabla^{2}}{2M}+ V_{\mathrm{opt}}(\mathbf{r},t)-E_{\mathrm{F}}+U(\mathbf{r},t).
\end{equation}
The Hartree potential is $U(\mathbf{r},t)=V \langle\hat{\psi}_{\alpha
}^{\dagger}(\mathbf{r},t)\hat{\psi}_{\alpha}(\mathbf{r},t) \rangle$, the pair
potential is $\Delta(\mathbf{r},t) = -V \langle\hat{\psi}_{\uparrow
}(\mathbf{r},t)\hat{\psi}_{\downarrow}(\mathbf{r},t) \rangle$, and
$u_{\mathbf{k}}(\mathbf{r},t)$ and $v_{\mathbf{k}}(\mathbf{r},t)$ are the
time-dependent \emph{quasi-particle} amplitudes. We denote the Fermi energy by
$E_{\mathrm{F}}=\hbar\omega_{\mathrm{F}}=\hbar^{2}k_{\mathrm{F}}^{2}/2M $, the
atom mass by $M$, and the strength of the collisional interaction between
fermions by $V$ ($V<0$). The Bragg field is $V_{\mathrm{opt}}(\mathbf{r}
,t)=A\cos(\mathbf{q}\cdot\mathbf{r}-\omega t)/2,$ where the wave vector
$\mathbf{q}$ is aligned with the $x$~axis.

The description of the collisional interaction in this problem requires a more
realistic collisional potential than a contact potential, since we find that
Bragg scattering is sensitive to the range of the potential.  However, this
sensitivity is weak for any realistic range, and we approximate the
collisional potential using a contact potential with strength $V$ and momentum
space cutoff $k_{\mathrm{c}}$, where $V$ and $k_{\mathrm{c}}$ are chosen to
give the correct scattering length and correct \emph{momentum space} range.
While the static BCS properties are insensitive to $k_{\mathrm{c}}$ \cite{Holland02}, the total number of Bragg scattered atoms depends on the value of the cutoff $k_{\mathrm{c}}$, and in fact the number of pairs scattered vanishes on setting $k_{\rm c}\rightarrow\infty$.  Thus, $k_{\mathrm{c}}$ must be chosen finite as above.  Full details will be given elsewhere \cite{Challis07}.

Details of our numerical method \cite{Challis07} are beyond the scope of this
paper. Briefly, the system ground state is found by solving iteratively for
the self-consistent fields in the time-independent form of Eq.~(\ref{bdgeqns}). The dynamic evolution of the gas is then determined by solving
Eq.~(\ref{bdgeqns}) using a fourth order Runge-Kutta method \cite{Ballagh00}.  We explicitly evolve approximately $40,000$ $\mathbf{k}$ modes in order to describe the dynamics of a Bragg scattered three-dimensional zero temperature homogeneous Fermi gas.

\begin{figure}[t]
\includegraphics[width=8cm]{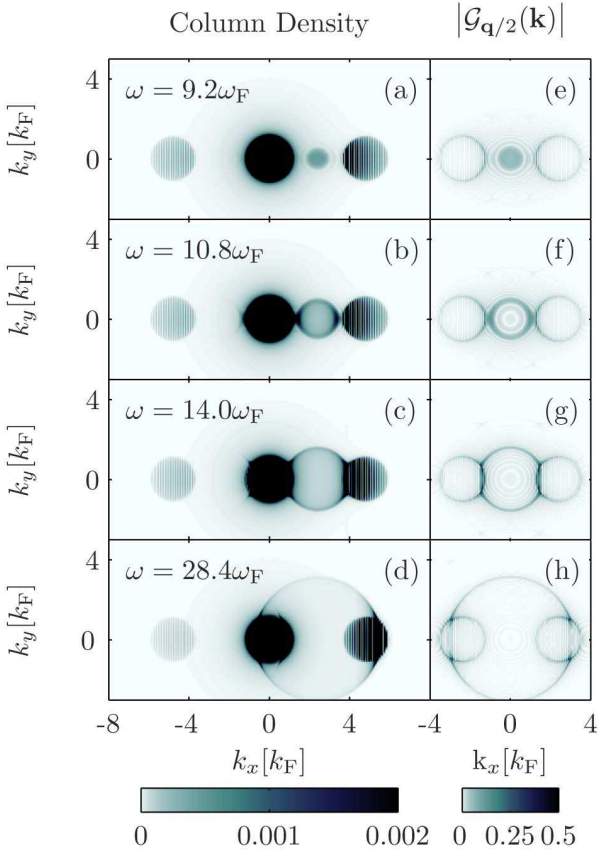}\caption{Column density, and pair
correlation function, of a Bragg scattered three-dimensional homogeneous Fermi
gas at zero temperature. The momentum distribution of the initial cloud is
saturated on the chosen scale in order to observe the scattered atoms.
Parameters are $A=1.80E_{\mathrm{F}},$ $q=4.80k_{\mathrm{F}},$ $t=8.22/\omega
_{\mathrm{F}},$ $k_{\mathrm{c}}=30k_{\mathrm{F}}$, $k_{\mathrm{F}}a=-0.427$
[i.e., $U(0)=-0.256E_{\mathrm{F}}$ and $\Delta(0)=0.049E_{\mathrm{F}}$], and
(a), (e) $\omega=9.2\omega_{\mathrm{F}},$ (b), (f) $\omega=10.8\omega
_{\mathrm{F}},$ (c), (g) $\omega=14.0\omega_{\mathrm{F}},$ and (d), (h)
$\omega=28.4\omega_{\mathrm{F}}.$}
\label{fig:vary_omega}
\end{figure}
Figure~\ref{fig:vary_omega}(a)-(d) shows the momentum space column
density of a Bragg scattered Fermi gas for different Bragg frequencies $\omega$. The column density is $\int n(\mathbf{k},t)dk_{z}$, with the number density at momentum $\hbar\mathbf{k}$ given by $n(\mathbf{k},t)=C\langle\hat{\phi}_{\alpha}^{\dagger}(\mathbf{k},t)\hat{\phi}_{\alpha}(\mathbf{k},t)\rangle$, where $\hat{\phi}_{\alpha}(\mathbf{k},t)=\int\hat{\psi}_{\alpha}(\mathbf{r},t)\exp
(-i\mathbf{k}\cdot\mathbf{r})d^{3}r/(L)^{3/2}$ is the field operator in momentum space, $C$ is a constant chosen such that $\int n(\mathbf{k},t)d^{3}k=1,$ and  $L^{3}$ is the computational volume.  The two spin states are scattered identically because of our choice of a spin preserving Bragg field.

The initial cloud, centered at $\mathbf{k}=\mathbf{0},$ has an approximate
width given by the modified Fermi wave vector $k_{\mathrm{F}}^{\prime}$,
defined by $\hbar^{2}k_{\mathrm{F}}^{\prime}{}^{2}/2M=E_{\mathrm{F}}-U(0)$
(e.g., \cite{Feder03}). In the presence of the Bragg field the atoms scatter
by two different mechanisms: (i) single-particle scattering, and (ii)
correlated-pair scattering.

In single-particle Bragg scattering an atom receives a momentum kick of $n\hbar\mathbf{q}$ and energy $n\hbar\omega,$ where $n$ is an integer.  A resonance condition selects primarily one value of $n$, and we consider $\omega$ in the range of first order Bragg scattering (e.g., \cite{Blakie00}), where $n=1$ is dominant.  This results in a cloud of atoms centered at $\mathbf{k}=\mathbf{q}$, as observed in Fig.~\ref{fig:vary_omega}(a)-(d).  A faintly visible cloud at $\mathbf{k}=-\mathbf{q}$ is also observed in Fig.~\ref{fig:vary_omega}(a)-(d) due to off resonant scattering into the $n=-1$ order.

Figure~\ref{fig:vary_omega}(a)-(d) also shows scattering of atoms into a
spherical shell centered at $\mathbf{k}=\mathbf{q}/2$. We refer to the
mechanism responsible for the atom shell as correlated-pair Bragg scattering.
Correlated-pair scattering has a frequency threshold denoted $\omega
_{\mathrm{thres}}$, at which atoms are scattered to $\mathbf{k}=\mathbf{q}/2$
[see Fig.~\ref{fig:vary_omega}(a)]. Above threshold [see
Fig.~\ref{fig:vary_omega}(b)-(d)], the atoms are scattered into a spherical
shell and the shell radius increases with Bragg frequency. Atoms scattered into
the shell come primarily from the Fermi surface, and are correlated about
$\mathbf{k}=\mathbf{q}/2$, as demonstrated by the pair correlation function
$\mathcal{G}_{\mathbf{q}/2}(\mathbf{k},t)=\langle\hat{\phi}_{\uparrow
}(\mathbf{q}/2+\mathbf{k},t)\hat{\phi}_{\downarrow}(\mathbf{q}/2-\mathbf{k},t)\rangle,$ shown in Fig.~\ref{fig:vary_omega}(e)-(h) for the $k_{z}=0$ plane
\footnote{The maximum possible correlation for a pair of atoms with momenta
$\pm\hbar\mathbf{k}$ is $|\mathcal{G}_{\mathbf{0}}(\mathbf{k},t)|=1/2.$}. As
well as the ring of correlation due to the scattered pairs, we also observe
two rings of radius $k_{\mathrm{F}}^{\prime}$ centered at $\mathbf{k}=\pm\mathbf{q}/2$, which represent correlation between an atom on the Fermi surface of the initial cloud, and an atom in the first order Bragg scattered cloud.

\begin{figure}[t]
\includegraphics[width=8cm]{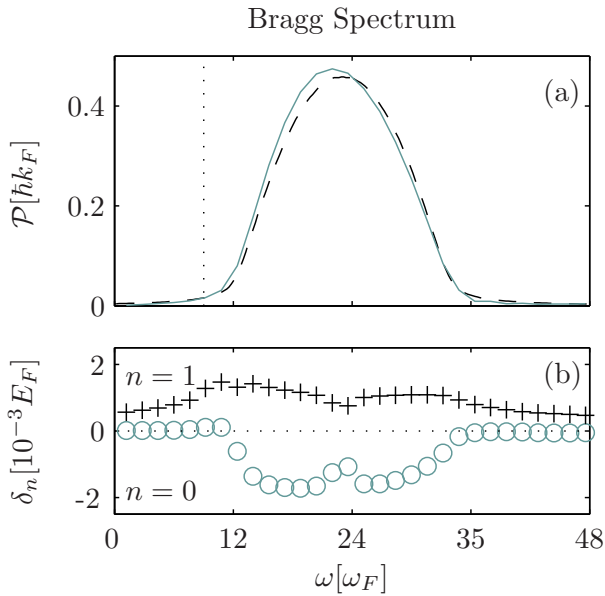}\caption{(a) Bragg spectrum of a
three-dimensional homogeneous Fermi gas at zero temperature. Parameters are
$A=1.80E_{\mathrm{F}},$ $q=4.80k_{\mathrm{F}},$ $t=8.22/\omega_{\mathrm{F}},$
$k_{\mathrm{c}}=30k_{\mathrm{F}}$, and (dashed) $k_{\mathrm{F}}a=0$ and
(solid) $k_{\mathrm{F}}a=-0.427.$  The non-interacting case is calculated with the initial momentum width of the gas modified to agree with the interacting case, i.e., $k_{\mathrm{F}}^{\prime}=1.12k_{\mathrm{F}}$.  The vertical dotted line indicates $\omega=\omega_{\mathrm{thres}}.$ (b) Relative change $\delta_{n}(t)=|\Delta_{n}(t)|-|\Delta_{n}(0)|$ of the pair potential Fourier components ($+$) $n=1$ and ($\circ$) $n=0$ [see Eq.~(\ref{pair_periodic})] for the case $k_{\mathrm{F}}a=-0.427$.}
\label{fig:spectra}
\end{figure}

We determine the Bragg spectrum of the degenerate Fermi gas by
calculating the momentum transfer per atom along the Bragg axis,
$\mathcal{P}(t)=\int[\hbar\mathbf{k}\cdot\hat{\mathbf{q}}]n(\mathbf{k},t)d^{3}k$, for a range of Bragg frequencies. The Bragg spectrum at zero
temperature is given in Fig.~\ref{fig:spectra}(a), and is dominated by the
broad single-particle resonance familiar from Bragg scattering of a
Bose-Einstein condensate (e.g., \cite{Blakie00}).  
The single-particle Bragg
resonance is due to two-photon scattering events that scatter atoms by the Bragg momentum $\hbar\mathbf{q}$. The resonance condition is well approximated using energy conservation arguments for non-interacting particles, and by considering an atom scattered from momentum $\hbar\mathbf{k}_{\mathrm{R}}$ to $\hbar(\mathbf{k}_{\mathrm{R}}+\mathbf{q})$ we obtain the resonance condition
\begin{equation}
\omega_{\mathrm{sp}}=\frac{\hbar}{2M}\left(  q^{2}+2\mathbf{k}_{\mathrm{R}}\cdot\mathbf{q}\right).
\end{equation}
The single-particle resonance [see Fig.~\ref{fig:spectra}(a)] is centered at
$\omega=\hbar q^{2}/2M,$ and its width is $\delta\omega\approx2\hbar
qk_{\mathrm{F}}^{\prime}/M+A/\hbar,$ where the first term accounts for the
momentum width of the initial cloud, and the second term is due to power
broadening (e.g., \cite{Blakie00}).

Correlated-pair Bragg scattering occurs on the red-detuned side of the
single-particle resonance, leading to a slight asymmetry in the Bragg spectrum
[see Fig.~\ref{fig:spectra}(a)]. The correlated scattering, which gives rise
to the distinctive spherical shell of atoms in momentum space [see Fig.~\ref{fig:vary_omega}(a)-(d)], depends
critically on the presence of Cooper pairs, and can not be understood by the usual single-particle scattering mechanism.
Correlated-pair scattering is associated with the formation of a moving
grating in the pair potential, which is well approximated by $\Delta
(\mathbf{r},t)\approx\Delta_{0}(t)+\Delta_{1}(t)\exp[i(\mathbf{q}\cdot\mathbf{r}-\omega t)].$ The pair potential coefficients $\Delta_{0}$ and
$\Delta_{1}$, after the Bragg pulse, are shown in Fig.~\ref{fig:spectra}(b),
where we observe that the homogeneous term $\Delta_{0}$ is depleted in the
region of the single-particle Bragg resonance, while the moving grating
amplitude $\Delta_{1}$, is created over a slightly more extended region.  Cooper pairs (of zero centre-of-mass momentum) scatter from the moving grating in the pair potential to centre-of-mass momentum $\hbar\mathbf{q}$. At threshold, [see Fig.~\ref{fig:vary_omega}(a)] the pair potential grating provides just sufficient energy to scatter each atom of a pair to a final momentum $\hbar\mathbf{q}/2$. Above threshold [see Fig.~\ref{fig:vary_omega}(b)-(d)], excess energy provided by the pair potential grating is distributed equally between the two atoms of a pair, and the individual momenta of the atoms become $\hbar(\mathbf{q}/2\pm
\mathbf{k}_{\mathrm{rel}})$. A resonance condition for correlated-pair Bragg
scattering can be obtained from energy conservation arguments to be \begin{equation}
\omega_{\mathrm{pair}}=\frac{\hbar}{M}\left(  \frac{q^{2}}{4}+k_{\mathrm{rel}}^{2}\right)  -\frac{\hbar k_{\mathrm{F}}^{\prime}{}^{2}}{M},
\label{omega_pair}
\end{equation}
which for $k_{\mathrm{rel}}=0$ gives the threshold frequency $\omega_{\mathrm{thres}}=\hbar q^{2}/4M-\hbar k_{\mathrm{F}}^{\prime}{}^{2}/M$.  Above threshold, the additional kinetic energy $\hbar
^{2}k_{\mathrm{rel}}^{2}/2M$ carried by each atom of a scattered pair is given by $\omega-\omega_{\mathrm{thres}}=\hbar
k_{\mathrm{rel}}^{2}/M$, as confirmed by our numerical calculations.

We can understand some important features of Bragg scattering of correlated
pairs with an analytic treatment. Due to the periodicity of the Bragg field,
the solutions of Eq.~(\ref{bdgeqns}) have the Bloch form, and can be expanded
as
\begin{align}
u_{\mathbf{k}}(\mathbf{r},t)  &  = e^{i\mathbf{k}\cdot\mathbf{r}} \sum_{n}
a_{n}^{\mathbf{k}}(t) e^{in(\mathbf{q}\cdot{\mathbf{r}}-\omega t)}\nonumber\\
v_{\mathbf{k}}(\mathbf{r},t)  &  = e^{i\mathbf{k}\cdot\mathbf{r}}\sum_{n}
b_{n}^{\mathbf{k}}(t) e^{in(\mathbf{q}\cdot{\mathbf{r}}-\omega t)},
\end{align}
where $n$ is an integer. The self-consistent potentials are periodic, with the
translational symmetry of the Bragg field, i.e.,
\begin{align}
U(\mathbf{r},t)  &  = \sum_{n} U_{n}(t)e^{in(\mathbf{q}\cdot\mathbf{r}-\omega
t)}\nonumber\\
\Delta(\mathbf{r},t)  &  = \sum_{n} \Delta_{n}(t) e^{in(\mathbf{q}%
\cdot\mathbf{r}-\omega t)}. \label{pair_periodic}
\end{align}
Evolution equations for the coeffcients $a_{n}^{\mathbf{k}}(t)$ and
$b_{n}^{\mathbf{k}}(t)$ can be derived from Eq.~(\ref{bdgeqns}) to be
\begin{align}
i \hbar\frac{da_{n}^{\mathbf{k}}(t)}{dt}  &  = \hbar\omega^{a}_{n}(\mathbf{k})
a_{n}^{\mathbf{k}}(t) + \frac{A}{4} \left[  a_{n+1}^{\mathbf{k}}(t) +
a_{n-1}^{\mathbf{k}}(t) \right] \label{evolvinga_int}\\
&  +\sum_{m} U_{m} (t)a_{n-m}^{\mathbf{k}}(t) + \sum_{m} \Delta_{m}
(t)b_{n-m}^{\mathbf{k}}(t),\nonumber
\end{align}
and
\begin{align}
i\hbar\frac{db_{n}^{\mathbf{k}}(t)}{dt}  &  = \hbar\omega^{b}_{n}(\mathbf{k})
b_{n}^{\mathbf{k}}(t) - \frac{A}{4} \left[  b_{n+1}^{\mathbf{k}}(t) +
b_{n-1}^{\mathbf{k}}(t) \right] \label{evolvingb_int}\\
&  -\sum_{m} U_{m}(t) b_{n-m}^{\mathbf{k}}(t) +\sum_{m} \Delta_{m}^{*}(t)
a_{n+m}^{\mathbf{k}}(t),\nonumber
\end{align}
where $\hbar\omega^{a,b}_{n} (\mathbf{k}) = \pm\left[  \hbar^{2}
(\mathbf{k}+n\mathbf{q})^{2}/2M-E_{\mathrm{F}} \right]  -n\hbar\omega.$
Initially the only non-zero coefficients are $a_{0}^{\mathbf{k}}$ (for
$|\mathbf{k}|\gtrsim k_{\mathrm{F}}^{\prime}$) and $b_{0}^{\mathbf{k}}$ (for
$|\mathbf{k}|\lesssim k_{\mathrm{F}}^{\prime}$). The mean-field coefficients
$U_{m}$ and $\Delta_{m}$ must be obtained self-consistently, in particular
\begin{equation}
\Delta_{m}(t)=-\bar{V}\sum_{\mathbf{k}}\sum_{n} a^{\mathbf{k}}_{n}
(t)b^{\mathbf{k}*}_{n-m}(t), \label{Delta_consistent}
\end{equation}
with $\bar{V}=T/(1-\alpha T),$ where $T$ is the low energy $T$-matrix and
$\alpha= Mk_{\mathrm{c}}/2\pi^{2}\hbar^{2}$ \cite{Holland02,Challis07}.

First order correlated-pair Bragg scattering is mediated by a moving grating
in the pair potential, arising due to terms $m=1$, $n=0,1$ in
Eq.~(\ref{Delta_consistent}) becoming non-zero. Those terms are seeded by
single-particle Bragg scattering events in which an atom on the Fermi surface
is scattered by momentum $\hbar\mathbf{q}$. It can be shown, from
Eqs.~(\ref{evolvinga_int}) and (\ref{evolvingb_int}), that single-particle
transitions generate coefficients $a_{1}^{\mathbf{k}}$ (for $|\mathbf{k}|\gtrsim k_{\mathrm{F}}^{\prime}$) and $b_{-1}^{\mathbf{k}}$ (for $|\mathbf{k}|\lesssim k_{\mathrm{F}}^{\prime}$). In the region of the Fermi
surface, $|\mathbf{k}|\approx k_{\mathrm{F}}^{\prime}$, $\Delta_{1}$ is
generated due to the formation of contributions $a_{1}^{\mathbf{k}}b_{0}^{\mathbf{k}\ast}$ and $a_{0}^{\mathbf{k}}b_{-1}^{\mathbf{k}\ast}$.
Physically this corresponds to seeding pair correlations about $\mathbf{k}=\mathbf{q}/2$. When an atom with momentum $\hbar \mathbf{k}_{\mathrm{R}}\approx \hbar k_{\mathrm{F}}^{\prime}\hat{\mathbf{k}}_{\rm {R}}$, is scattered by a single-particle transition to $\hbar(\mathbf{k}_{\mathrm{R}}+\mathbf{q}),$ it
remains correlated with its unscattered pair at momentum $-\hbar \mathbf{k}_{\mathrm{R}}$.  This reduces the system pairing about $\mathbf{k}=\mathbf{0}$ ($\Delta_{0}$), and generates a pair correlation about
$\mathbf{k}=\mathbf{q}/2$ ($\Delta_{1}$) [see Fig.~\ref{fig:spectra}(b)].  This correlation seeding can also be observed in $\mathcal{G}_{\mathbf{q}/2}(\mathbf{k},t)$ [see Fig.~\ref{fig:vary_omega}(e)-(h)], where a point on the left most circle represents correlation between an atom on the Fermi surface of the initial cloud and its pair which has been scattered by $\mathbf{q}$.

Following the initiation of the pair potential grating, its subsequent
development can be understood in terms of a truncated version of
Eqs.~(\ref{evolvinga_int}) and (\ref{evolvingb_int}), i.e.,
\begin{equation}
i\hbar\frac{d}{dt}\left[
\begin{array}
[c]{c}
a_{0}^{\mathbf{k}}(t)\\
b_{-1}^{\mathbf{k}}(t)
\end{array}
\right]  =\left[
\begin{array}
[c]{cc}
\epsilon_{0}^{a}(\mathbf{k}) & \Delta_{1}(t)\\
\Delta_{1}^{\ast}(t) & \epsilon_{-1}^{b}(\mathbf{k})
\end{array}
\right]  \left[
\begin{array}
[c]{c}
a_{0}^{\mathbf{k}}(t)\\
b_{-1}^{\mathbf{k}}(t)
\end{array}
\right]  ,
\label{d1_eqs}
\end{equation}
which is appropriate for describing the scattered pairs. In Eq.~(\ref{d1_eqs}), $\epsilon_{n}^{a,b}(\mathbf{k})=\hbar\omega_{n}^{a,b}(\mathbf{k})\pm U_{0}$, and
$\Delta_{1}(t)=-\bar{V}\sum_{\mathbf{k}}a_{0}^{\mathbf{k}}(t)b_{-1}^{\mathbf{k}\ast}(t)$. The term $a_{0}^{\mathbf{k}}b_{-1}^{\mathbf{k}\ast}$ only becomes significant if $\epsilon=\epsilon_{0}^{a}(\mathbf{k})-\epsilon_{-1}^{b}(\mathbf{k})\approx0,$ i.e., when correlated scattering transitions conserve energy ($\omega\geq\omega_{\mathrm{thres}}$). At threshold, the summands in $\Delta_{1}$ have a stationary phase, leading to enhancement of the grating amplitude $\Delta_{1}$ [see Fig.~\ref{fig:spectra}(b)]. The thickness $\delta k$ of the spherical shell of scattered pairs can be estimated by assuming a frequency width $\Gamma$, determined by the Bragg pulse length ($\Gamma\approx\pi/t$), and setting $\delta\epsilon=\hbar\Gamma,$
to find that
\begin{equation}
\delta k\approx\sqrt{\frac{\pi M}{\hbar t}+k_{\mathrm{rel}}^{2}}-k_{\mathrm{rel}}.
\end{equation}

We have investigated the dependence of correlated-pair Bragg scattering on a
range of system parameters.  In Fig.~\ref{fig:vary_omega}(b), $\sim0.2\%$ of the atoms are scattered by correlated-pair Bragg scattering, and the number of scattered pairs grows linearly with the length of the Bragg pulse (until $t \sim 70/\omega_{\rm F}$).  Over the range $-0.18 \geq k_{\mathrm{F}}a\geq-0.69$, the number of pairs scattered increases quadratically with $\Delta(0)$ indicating the coherent nature of the scattering process.  For $k_{\mathrm{F}}a=-0.689$ [all other parameters as per Fig.~\ref{fig:vary_omega}(b)] there are $\sim6\%$ scattered pairs.  However, we emphasize that for $|k_{\mathrm{F}}a|\gtrsim1$ the mean-field approach may not be quantitatively accurate (e.g., \cite{Engelbrecht97,Combescot06}).  The number of correlated pairs scattered can be further increased by enhancing the single-particle scattering processes that seed the pair potential grating, either by increasing the Bragg field strength $A$, or by reducing the Bragg wave vector $\mathbf{q}$ (to make the seeding more resonant).

In conclusion, we have calculated solutions of the time-dependent Bogoliubov
de Gennes equations for a zero temperature homogeneous three-dimensional Bragg
scattered Fermi gas, in the regime where the momentum transfer is well outside
the Fermi surface. We predict Bragg scattering of correlated atom pairs, which
has a distinctive signature in momentum space, namely a spherical shell of atoms
centered at half the usual Bragg momentum transfer. Correlated-pair Bragg
scattering occurs via a Bragg grating formed in the pair potential, and has a
well defined frequency threshold on the red-detuned side of the familiar
single-particle Bragg resonance. We have developed an analytic model that
explains the mechanism by which the pair potential grating is generated, and
observe that the number of scattered pairs is proportional to the square of
the initial pairing field.

\begin{acknowledgments}
This work was supported by Marsden Fund UOO0509 and the Tertiary Education
Commision (TAD 884).
\end{acknowledgments}

\end{document}